\documentclass[journal=jpccck,manuscript=article]{achemso}
\usepackage[english]{babel}
\usepackage{graphicx}
\usepackage{epsfig}
\usepackage{color}
\usepackage{amsfonts,amssymb,amsmath}
\newcommand{\ISM}[0]{{
 Universit\'e de Bordeaux, ISM, UMR 5255, F-33400 Talence, France}}
\newcommand{\ISMCNRS}[0]{{
 CNRS, ISM, UMR 5255, F-33400 Talence, France}}

\author{L. Bonnet}
\email{claude-laurent.bonnet@u-bordeaux.fr}
 \affiliation{\ISM}
 \alsoaffiliation{\ISMCNRS}

\title{
Semiclassical initial value theory of rotationally inelastic scattering: Some remarks on the phase index in the interaction picture
}


\SectionNumbersOn 

\begin{document}

\begin{abstract}
This paper deals with the treatment of quantum interferences in the semiclassical initial value theory of rotationally inelastic scattering 
in the interaction picture [C. W. McCurdy and W. H. Miller, J. Chem. Phys. \textbf{67}, 463 (1977)]. It is shown that substituting the original 
phase index by a new one extends the range of applicability of the theory. The resulting predictions are in close agreement 
with exact quantum scattering results for a model of atom-rigid diatom collision involving strong interferences. 
The developments are performed within the framework of the planar rotor model, but they are readily
applicable to the three-dimensional case. 
\end{abstract}

\section{Introduction}
\label{I}

Numerous calculations performed over the last two decades\cite{Aoiz1,Aoiz-1,Banares,Kim,Bowman0,Braunstein1,Aoiz0,GB6,SBG,Gabor3,ZhangGuo,NGB,
GB4,Wilmer,Gabor0,Gyorgy,Braunstein2,Fu1,Dario0,Gabor-1,CRChimie,Cao,Rob,LBJE,Note1} have shown that the classical trajectory method\cite{Karplus1,
Porter,Sewell,Aoiz1,Hase} generally allows 
the nearly quantitative description of the dynamics and the kinetics of molecular processes relevant to atmospherical and interstellar chemistry.
These calculations may require quantum corrections to improve their accuracy, like the use of Wigner distributions,\cite{BrownHeller,Wilmer} 
one-dimensional tunneling probabilities,\cite{Schatz,Makri,Thompson,Takatsuka,Coronado,Brandao} Gaussian binning,\cite{GB4,GB6} surface 
hopping,\cite{Nakamura1,Barbatti,Yonehara,Tully,Nakamura2} etc., but contrary to semiclassical calculations,\cite{Bill1,Bill2,Marcus,SemiClass3,
SemiClass4,SemiClass2,Heller1,Campo1,Shala,Rost1,ElranKay,Saraceno,BC1,BC2} they do not assign probability 
amplitudes and phases to classical paths. In this regard, the classical trajectory method mostly ignores the wave character of nuclear motions
involved in molecular processes.  

Sometimes, however, these calculations are ``too classical'' and lack of realism. This is typically the case for rotationally inelastic atom-diatom 
collisions, of great importance in interstellar chemistry\cite{Lique} and stereodynamics.\cite{Stolte1,Stolte2,Chandler1,Suzuki1,QQT,Brouard0,
Groenenboom,Brouard1,Brouard2} 
Unlike chemical reactions, inelastic collisions may involve strong interference effects that are only partially (if not at all) quenched by 
the summation over total and orbital (or helicity) quantum numbers involved in the calculation of integral cross sections (ICS). This is typically
the case for collisions of near-homonuclear (almost symmetrical) molecules such as NO with noble gasses.\cite{Stolte1,Stolte2,Chandler1,Suzuki1,
QQT,Brouard0,Groenenboom,Brouard1,Brouard2}
For such processes, interference effects show up in final rotational state distributions, or in the steric 
asymmetry measuring the dependence of the previous distributions on the initial orientation of the diatom with respect to the atom (quantum 
expectations oscillate about classical ones). 

In an illuminating analysis, McCurdy and Miller\cite{McCurdy} showed within a planar model of atom-diatom inelastic collision that 
the semiclassical theory of molecular collision\cite{Bill1,Bill2,Marcus,SemiClass3,
SemiClass4,SemiClass2,Heller1,Campo1,Shala,Rost1,ElranKay,Saraceno,BC1,BC2} 
not only reproduces the previous interference features, but also provides deep insight into their physical origin. The 
primary reason for this success is that, as previously stated, semiclassical methods assign probability amplitudes 
and phases to classical trajectories and make them interfere, respecting thereby the quantum principle of superposition. 
The accuracy with which one makes these paths interfere is the subject of this report.

It should be noted that for atom-diatom inelastic collisions, exact quantum scattering (EQS) calculations are nearly 
routine nowadays.\cite{ArDal,Millard,Note3}
Moreover, the promizing implementation of the mixed quantum/classical theory by Semenov and Babikov\cite{Bab1,Bab2,Bab3,Bab4} may also provide 
accurate results at a lower computational cost.
These benchmark calculations, however, are often too complex to provide insight into the physics underlying interference effects. 
The main interest of the semiclassical approach of inelastic collisions is thus its explicative power. 

One may thus wonder why this approach has never been used to reproduce and analyze the state-of-the-art stereodynamics measurements performed 
over the last two decades.\cite{Stolte1,Stolte2,Chandler1,Suzuki1,QQT,Brouard0,Groenenboom,Brouard1,Brouard2}
This is likely due to major and somehow discouraging numerical difficulties encountered in the application of semiclassical scattering methods, 
especially when the dynamics involves trapped trajectories.\cite{Saraceno} 
However, the processes under scrutiny in stereodynamical studies\cite{Stolte1,Stolte2,
Chandler1,Suzuki1,QQT,Brouard0,Groenenboom,Brouard1,Brouard2} involve a single rebound mechanism and hence, no trapping. For such type of
encounters, at least two previous works by Miller\cite{Bill3} and Campolieti and Brumer\cite{Campo1} suggest that realistic 
(full-dimensional) semiclassical calculations should be feasible. 

In principle, accurate semiclassical (SC) predictions are expected to be obtained within the \emph{initial value representation} (IVR) 
discovered by Miller,\cite{Bill2}
and first applied to vibrationally inelastic collisions. This method is developed within the \emph{interaction picture}, as we shall see further 
below. The major interests of the SCIVR approach are twofold. First, it allows to go round the root-search issue of \emph{classical S-matrix theory} 
(CSMT),\cite{Bill1,Bill2,Marcus} mother of all semiclassical methods of molecular scattering. Second, 
it is able to predict quantum mechanically allowed 
transitions that are classically forbidden, contrary to CSMT. 
For rotationally inelastic collisions, Miller's SCIVR $S$-matrix elements 
are given by Eq.\;(3.5) in Ref.~\cite{McCurdy} for the planar rotor, and Eq.\;(12) in Ref.~\cite{Campo1} for the three-dimensional 
rotor. The goal of this work is to increase the range of applicability of these expressions by considering a phase index different from 
the original one. The developments are performed within the framework of the planar rotor model, but they can be straightforwardly 
extended to the three-dimensional case. 

The paper is laid out as follows. The previously outlined SCIVR method\cite{McCurdy} is presented and applied to a model of atom-planar rotor 
collision in Sec.~\ref{II}. Three coupling strengths between translational and rotational motions are considered. For the two lowest ones, 
SCIVR predictions are in close agreement with EQS results, but for the strongest one, clear disagreement is found. 
In order to shed light on this finding, the conditions of validity of the SCIVR method are analyzed in Sec.~\ref{III}. 
In particular, we show that the latter does not necessarily lead to CSMT in the classical limit. 
This inconsistency is removed in Sec.~\ref{IV} by modifying the phase index. SCIVR predictions are 
then in close agreement with EQS results for the three coupling strengths. A technical discussion follows in Sec.~\ref{V} and 
Sec.~\ref{VI} concludes.

\section{SCIVR theory in the interaction picture}
\label{II}

We consider, within a fixed-plane of the laboratory frame, the collision between an atom and a rigid diatom rotating in the previous plane. 
Moreover, both the atom and the center-of-mass of the rotor are supposed to move on a fixed line of the plane throughout 
the collision (see Fig. 1 in Ref.~\cite{BC1}). Detailed discussions of the classical, 
semiclassical and quantum dynamics of this collisional system can be found elsewhere.\cite{SemiClass4,BC1,BC2} Calling 
$R$ the distance between the atom and the center-of-mass of the rotor, $\phi$ the Jacobi angle  
and $P$ and $J$ their respective conjugate momenta, the classical Hamiltonian of the system is given by 
\begin{equation}
 H = H_0 + V(R,\phi)
 \label{1a}
\end{equation}
where
\begin{equation}
 H_0 = \frac{P^2}{2\mu}+\frac{J^2}{2I}
 \label{1b}
\end{equation}
is the unperturbed Hamiltonian. 
$\mu$ is the reduced mass of the atom-diatom system. $I$ is the moment of inertia of the diatom, given by 
\begin{equation}
 I=m r^2
 \label{1c}
\end{equation}
were $m$ is the reduced mass of the diatom and $r$ its bond length. $V(R,\phi)$ is the interaction potential. 
In addition to $\phi$, we introduce the shifted Jacobi angle
\begin{equation}  
\tilde{\phi} = \phi- \frac{\mu RJ}{PI}.
 \label{2}
\end{equation} 
$\mu R/P$ is the time to go from 0 to $R$ when ignoring the interaction between the atom and the diatom, i.e., when assuming that 
the dynamics is governed by $H_0$ instead of $H$. Moreover, $J/I$ is the angular velocity of the rotor. $\mu RJ/(PI)$ is thus the 
variation of $\phi$ when going from 0 to $R$ if the Hamiltonian of the system is $H_0$. Hence, $\tilde{\phi}$ results from 
making $\phi$ evolve forward in time according to $H$ and then backward in time according to $H_0$. This evolution is 
analogous to that of a quantum state in the interaction picture.\cite{Skodje} Note that $\tilde{\phi}$ is a constant of motion 
in the asymptotic channel where $V(R,\phi)$ is zero. 

Let $S_{j_2j_1}(E)$ be the probability amplitude to go from the initial rotational state $j_1$ to the final state $j_2$ at the 
total energy $E$. The set of trajectories used further below to calculate this element is defined as follows.
They are started at $R_1$, large enough for $V(R,\phi)$ to be negligible, with $H = E = H_0$ (see Eq.~\eqref{1a}). 
Moreover, the initial angular momentum $J_1$ of the rotor is kept at $\hbar j_1$. From Eq.~\eqref{1b}, we thus have
\begin{equation}
 P_1 = -\left[2\mu\left(H_0-\frac{J_1^2}{2I}\right)\right]^{1/2} = -\left[2\mu\left(E-\frac{\hbar^2j_1^2}{2I}\right)\right]^{1/2}.
 \label{3a}
\end{equation}
The initial shifted angle $\tilde{\phi}_1$ can take any value within the range $[0,2\pi]$. The resulting trajectories cross the interaction 
region and eventually come back to the asymptotic channel. They are finally stopped at $R_2$, large enough for $V(R,\phi)$ to be negligible. 
The values of $\tilde{\phi}$ and $J$ at $R_2$ are denoted $\tilde{\phi}_2$ and $J_2$, respectively. Since both $\tilde{\phi}$ and $J$ are 
constants of motion in the asymptotic channel, $\tilde{\phi}_2$ and $J_2$ are their final values. From Eq.~\eqref{1b}, we have 
\begin{equation}
 P_2 = \left[2\mu\left(E-\frac{J_2^2}{2I}\right)\right]^{1/2}.
 \label{3b}
\end{equation}
We note from Eq.~\eqref{2} and the left equality of Eq.~\eqref{3a} that 
\begin{equation}
 \frac{\partial X_2}{\partial\tilde{\phi}_1}\Bigg\vert_{J_1}=\frac{\partial X_2}{\partial{\phi}_1}\Bigg\vert_{J_1}
 \label{3c}
\end{equation}
with $X_2$ equal $J_2$ or $\tilde{\phi}_2$. This identity will be useful in the following.

Miller's SCIVR expression of $S_{j_2j_1}(E)$ reads
\begin{equation}
 S_{j_2j_1}(E) = \frac{1}{2\pi i} \int_{0}^{2\pi} d\tilde{\phi}_1 
 \;\left(
 \frac{\partial\tilde{\phi}_2}{\partial\tilde{\phi}_1}\Bigg\vert_{J_1}
 \right)^{1/2}
 e^{i\Phi/\hbar}
 \label{3}
\end{equation}
with
\begin{equation}
 \Phi
 = \left(J_2-\hbar j_2\right)\tilde{\phi}_2+\Omega
 \label{4}
\end{equation}
and
\begin{equation}
 \Omega
 = - \int_0^t d\tau \left(R\dot{P}+\phi \dot{J}\right)
 \label{4a}
\end{equation}
(see Eq. (3.5) in Ref.~\cite{McCurdy}; the only difference is that in the present Eq.~\eqref{3}, an overall and arbitrary phase factor $1/i$ has been 
added for consistency with previous developments,\cite{BC1} and $\hbar$ is not kept at 1). The partial derivative 
$\frac{\partial\tilde{\phi}_2}{\partial\tilde{\phi}_1}\big\vert_{J_1}$ is deduced from the set of trajectories previously introduced
and a second batch of nearby paths starting with the same initial conditions but slightly different values of $\tilde{\phi}_1$. 
Eq.~\eqref{3} can be rewritten as
\begin{equation}
 S_{j_2j_1}(E) = \frac{1}{2\pi i} \int_{0}^{2\pi} d\tilde{\phi}_1 
 \;\left|
 \frac{\partial\tilde{\phi}_2}{\partial\tilde{\phi}_1}\Bigg\vert_{J_1}
 \right|^{1/2}
 e^{i\left(\Phi/\hbar-\pi\nu/2\right)}.
 \label{5}
\end{equation}
The phase index $\nu$ is equal to 0 if $\frac{\partial\tilde{\phi}_2}{\partial\tilde{\phi}_1}\big\vert_{J_1}$ is positive. In the contrary case,
$\nu$ is equal to $\pm 1$. The two signs are considered since the form of the pre-exponential factor in Eq.~\eqref{3} 
does not allow to decide which branch of the square root should be chosen. The calculations performed by means of this first index
will be called SC-I$^-$ and SC-I$^+$ for $\nu = (0,-1)$ and $(0,+1)$, respectively. 

\begin{figure}[!h]

\centering{\includegraphics[width=27cm]{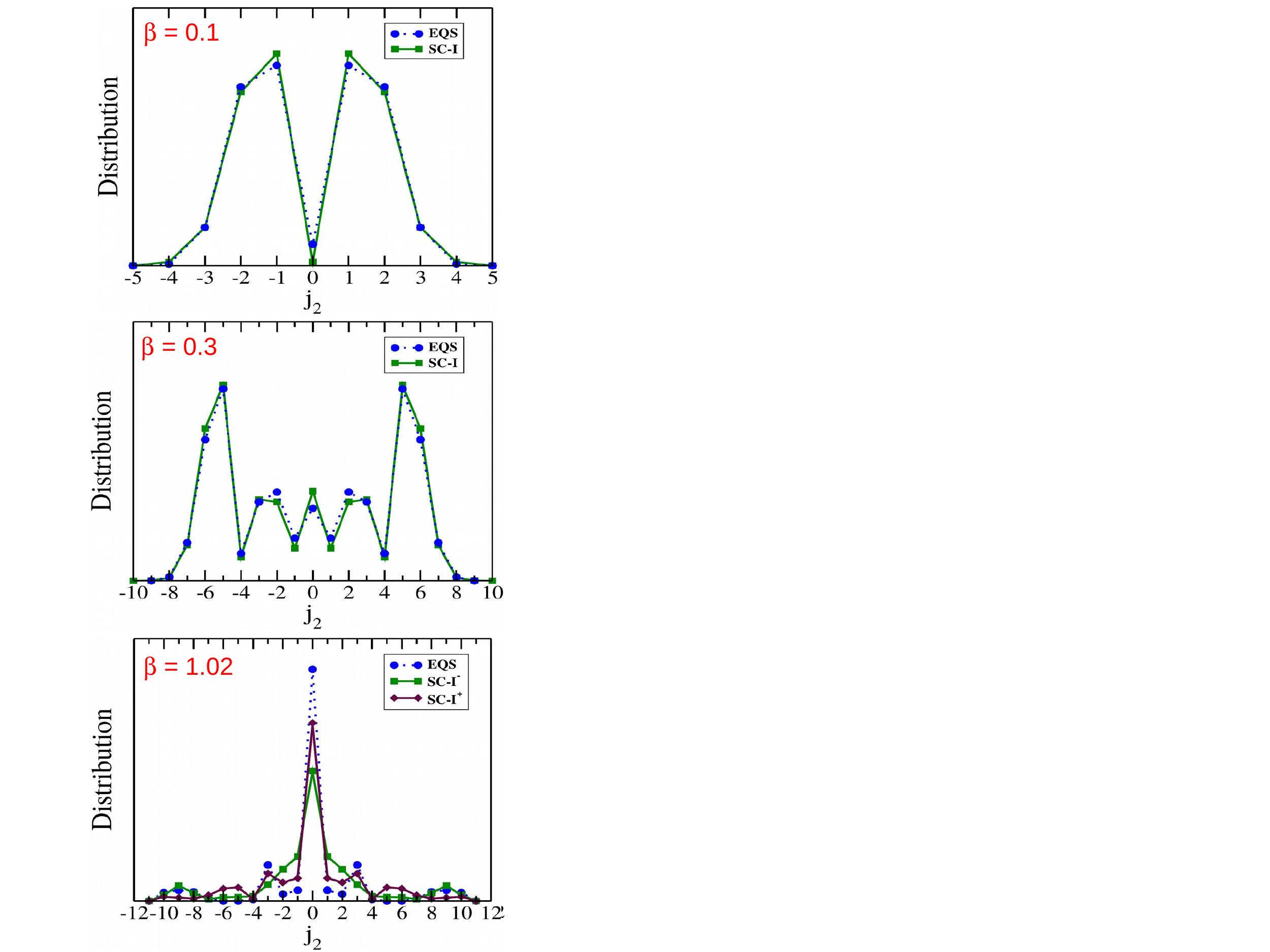}}

\caption{Rotational state distributions found from exact quantum scattering (EQS) calculations (blue circles connected by dotted segments)
and semiclassical calculations according to Eq.~\eqref{5} (green squares or brown diamonds connected by solid segments). Three different values of 
the $\beta$ parameter are considered in Eq.~\eqref{6}. The acronyms SC-I, SC-I$^-$ and SC-I$^+$ correspond to different
definitions of the index $\nu$ in Eq.~\eqref{5} (see text).}

\label{fig:1}

\end{figure}

The model planar collision used to check the validity of Eq.~\eqref{3} is governed by $H$ with\cite{BC1}
\begin{equation}
 V(R,\phi) = exp\left[{-\alpha \left(R - \beta cos\phi\right)}\right].
 \label{6}
\end{equation} 
While $\alpha$ is fixed at 2 \AA$^{-1}$, $\beta$ is taken at the three different values 0.1, 0.3, and 1.02 \AA\;corresponding to increasing 
couplings between the $R$ and $\phi$ coordinates in the interaction region. E is kept at 0.5 eV, $j_1$ at 0, $\mu$ at 2/3 amu, $m$ at 1/2 amu and
$r$ at 1 \AA. $R_1$ and $R_2$ are both taken at 4 \AA, beyond which $V(R,\phi)$ is negligibly small and the integrand 
in Eq.~\eqref{5} is a constant of motion. 
The collisional systems resulting from the previous parameters involve strong interferences, as shown by the exact quantum 
state distributions $P_{j_2j_1}(E)=|S_{j_2j_1}(E)|^2$
displayed in Fig.~\ref{fig:1} (blue circles connected by dotted segments; see Ref.~\cite{BC1} for some details on their calculations). 
Since $J_1=0$, we note from Eq.~\eqref{2} that $\tilde{\phi}_1={\phi}_1$. 

\begin{figure}[!h]

\centering{\includegraphics[width=27cm]{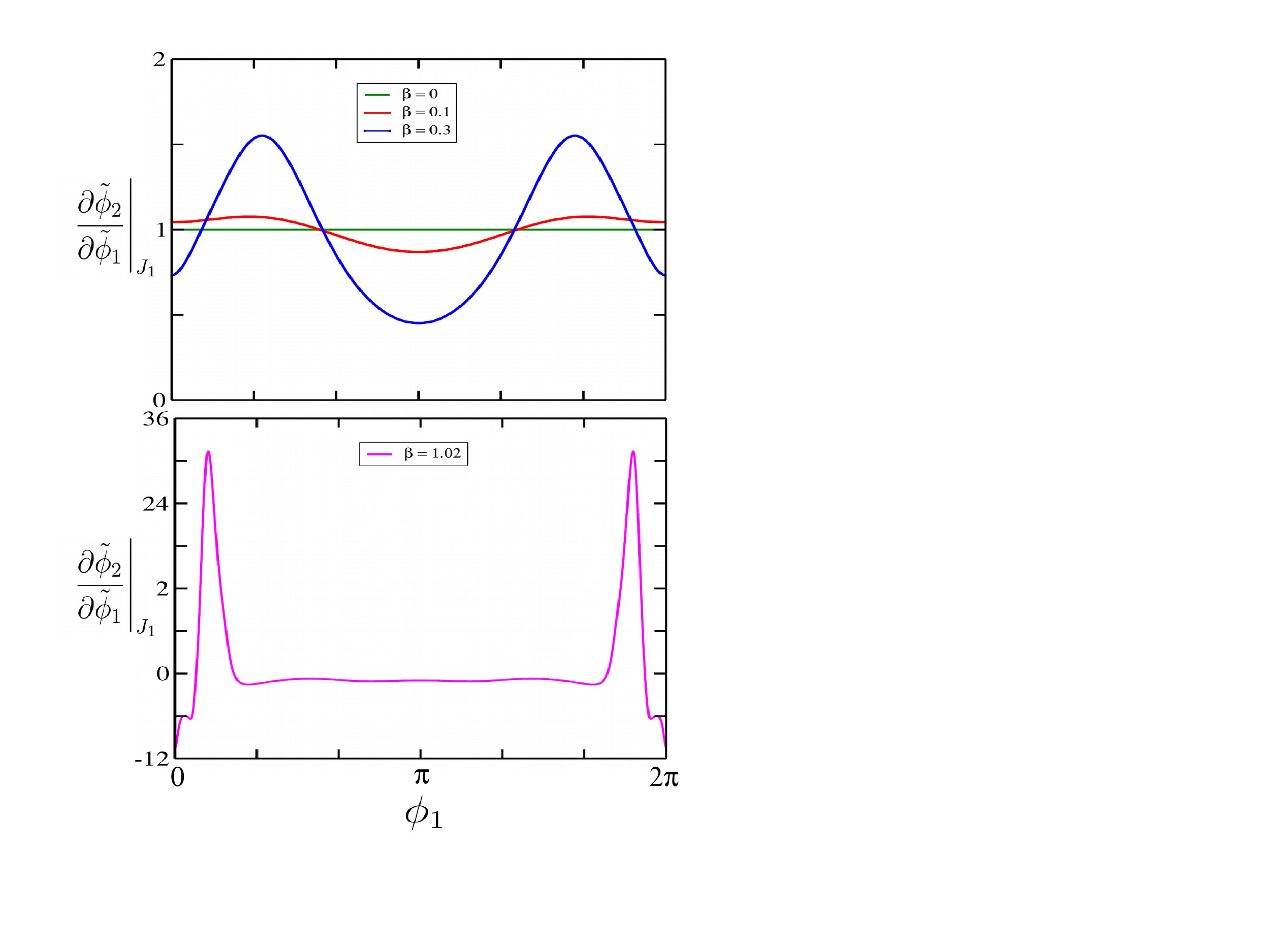}}

\caption{$\frac{\partial\tilde{\phi}_2}{\partial\tilde{\phi}_1}\big\vert_{J_1}$ in terms of ${\phi}_1$ ($=\tilde{\phi}_1$) 
for $\beta$ equal 0, 0.1 and 0.3 (upper panel) and 1.02 (lower panel).}

\label{fig:2}

\end{figure}

For $\beta=0$, there is no coupling between $R$ and $\phi$ and $J$ keeps constantly 
equal to $\hbar j_1$ during the collision. Calling $t(R_1,R_2)$ the time to go from $R_1$ to the interaction region and back to $R_2$, we have 
\begin{equation}
 \phi_2=\phi_1+\frac{\hbar j_1}{I}\;t(R_1,R_2).
 \label{5a}
\end{equation}
Using Eq.~\eqref{2}, we thus arrive at
\begin{equation}  
\tilde{\phi_2} = \phi_1+\frac{\hbar j_1}{I}\;t(R_1,R_2)-\frac{\mu R_2 \hbar j_1}{P_1 I}
 \label{5b}
\end{equation}
with $P_1$ given by Eq.~\eqref{3a}. From Eqs.~\eqref{3c} and~\eqref{5b}, and the fact that $t(R_1,R_2)$ does not depend on $\phi_1$, 
we obtain\begin{equation}
 \frac{\partial\tilde{\phi}_2}{\partial\tilde{\phi}_1}\Bigg\vert_{J_1}=1.
 \label{5c}
\end{equation}
$\frac{\partial\tilde{\phi}_2}{\partial\tilde{\phi}_1}\big\vert_{J_1}$ is represented in Fig.~\ref{fig:2} for $\beta=0$ and the three values
previously considered. 
For $\beta$ equal 0.1, the coupling is small and $\frac{\partial\tilde{\phi}_2}{\partial\tilde{\phi}_1}\big\vert_{J_1}$ slightly oscillates 
around 1.
For $\beta$ equal 0.3, the coupling is stronger, thus leading to oscillations around 1 of larger amplitude. 
In both cases, however, 
$\frac{\partial\tilde{\phi}_2}{\partial\tilde{\phi}_1}\big\vert_{J_1}$ is found to be positive.
Hence, the SC-I$^-$ and SC-I$^+$ approaches lead to the same results, simply labeled SC-I 
(green squares connected by solid segments in the upper and middle panels of Fig.~\ref{fig:1}). 
As a matter of fact, the agreement between quantum and semiclassical predictions is very satisfying. 

On the other hand, for $\beta$ equal 1.02, the strength of the coupling is such that 
$\frac{\partial\tilde{\phi}_2}{\partial\tilde{\phi}_1}\big\vert_{J_1}$ 
can take both signs. The SC-I$^-$ and SC-I$^+$ approaches then lead to different results (green squares and brown diamonds connected by solid 
segments in the lower panel of Fig.~\ref{fig:1}). Both of them appear to be in quantitative disagreement with EQS results as compared to the 
two previous cases. In Sec.~\ref{III}, we determine the origin of this problem and in Sec.~\ref{IV}, we show how to fix it.

\section{Condition of validity of Eq.~\eqref{5}}
\label{III}

Let us momentarily abandon Eq.~\eqref{5} to concentrate on CSMT,\cite{Bill1,Bill2,Marcus} 
the ``most classical'' semiclassical approach of molecular collisions. The CSMT expression of $S_{j_2j_1}(E)$ reads
\begin{equation}
 S_{j_2j_1}(E) = \sum_{k} 
 \;
 \left(
 \frac{2\pi i}{\hbar}
 \left|
 \frac{\partial J_2}{\partial\tilde{\phi}_1}\Bigg\vert_{J_1}^k
 \right|
 \right)^{-1/2}
 e^{i\left(\Omega_k/\hbar-\pi\eta_k/2
 \right)}
 \label{7}
\end{equation}
(we assume here that $i^{1/2}=e^{i\pi/4}$).  
The sum is over the discrete set of trajectories starting from $R_1$ with $P_1$ given by the right-hand-side (RHS) of Eq.~\eqref{3a} 
and reaching $R_2$ with $J_2=\hbar j_2$. The new quantity 
appearing here is $\eta_k$, the Maslov index\cite{Littlejohn1} of the classical Green function, equal here to the number of caustics touched 
by the $k^{th}$ trajectory. These are located as follows; a second path is considered, starting from the same point 
$(R_1,\phi_1^k)$ as the $k^{th}$ trajectory, but with initial momenta differing by an infinitely small amount from those of the
$k^{th}$ trajectory (with the constraint $H_0=E$ still satisfied). The $k^{th}$ trajectory touches a caustic the instant it is crossed by 
the second path in the $(R,\phi)$ plane. The practical calculation of $\eta_k$ is detailed in section II.A.3 of ref.\cite{BC1}. 
In the same work, two trajectories involving one and two caustics are schematically represented in Figs. 5 and 6, respectively. 
They correspond to the blue lines. The green lines in the lower panels are the second paths, crossing the blue lines once (Fig. 5) 
or twice (Fig. 6). It is demonstrated below these figures that $\frac{\partial J_2}{\partial\tilde{\phi}_1}\big\vert_{J_1}$ is 
positive (negative) for one (two) caustic(s). This finding will be useful in the following.
A detailed derivation of Eq.~\eqref{7} from first principles can be found in Refs.~\cite{BC1,BC2}. Eq.~\eqref{7} 
above is identical to Eq. (103) in Ref.~\cite{BC1}, where $g$ is in fact equal to 1~\cite{BC2} 
and $\frac{\partial J_2}{\partial{\phi}_1}\big\vert_{J_1}^k$ replaces $\frac{\partial J_2}{\partial\tilde{\phi}_1}\big\vert_{J_1}^k$,
a substitution justified by Eq.~\eqref{3c}.

CSMT is the theory toward which any kind of SCIVR approach is expected to converge in the classical limit.\cite{Bill2,Campo1,ElranKay,BC1} 
Hence, one should recover Eq.~\eqref{7} from Eq.~\eqref{5} when making $\hbar$ tend to 0 in the latter. 
Note, however, that the resulting expression may differ 
from the RHS of Eq.~\eqref{7} by an irrelevant phase factor, noted $e^{i\pi l/2}$ without loss of generality, 
disappearing when considering $|S_{j_2j_1}(E)|^2$. 
In order to introduce the standard result of asymptotic analysis necessary to perform the passage from Eq.~\eqref{5} to Eq.~\eqref{7}, 
consider two functions $f$ and $g$, the former involving $N$ stationary points $(x_1,...,x_N)$. For an infinitely small parameter $s$ 
(we take it positive), it can be shown that
\begin{equation}
 \int dx \; g(x) e^{if(x)/s} = \sum_{k=1}^N g(x_k) \left(\frac{2\pi is}{|f''(x_k)|}\right)^{1/2} 
 e^{i[f(x_k)/s-\pi\chi_k/2]}
 \label{8}
\end{equation}
where $\chi_k$ is 0 if $f''(x_k)$ is positive, 1 otherwise.\cite{Gutzwiller} When applied to the case where $s$ is small but 
not negligible with respect to the $|f''(x_k)|$'s, Eq.~\eqref{8} is known as the \emph{stationary phase approximation} (SPA). 
We now use it to integrate the RHS of Eq.~\eqref{5} over $\tilde{\phi}_1$. 
First of all, we rewrite Eq.~\eqref{4a} as
\begin{equation}
 \Omega
 = - \int_{P_1}^{P_2} R dP - \int_{\hbar j_1}^{J_2} \phi dJ.
 \label{4b}
\end{equation}
Since only the upper bounds $P_2$ and $J_2$ depend on $\tilde{\phi}_1$, we have
\begin{equation}
 \frac{\partial\Omega}{\partial\tilde{\phi}_1}\Bigg\vert_{J_1}
 = - R_2 \frac{\partial P_2}{\partial\tilde{\phi}_1}\Bigg\vert_{J_1} - \phi_2 \frac{\partial J_2}{\partial\tilde{\phi}_1}\Bigg\vert_{J_1}.
 \label{4c}
\end{equation}
Using the fact that $P_2$ and $J_2$ satisfy Eq.~\eqref{1b} (with $H_0=E$) and using Eq.~\eqref{2} allows to rewrite Eq.~\eqref{4c} as
\begin{equation}
 \frac{\partial\Omega}{\partial\tilde{\phi}_1}\Bigg\vert_{J_1}
 = - \tilde{\phi}_2 \frac{\partial J_2}{\partial\tilde{\phi}_1}\Bigg\vert_{J_1}, 
 \label{4d}
\end{equation}
which, together with Eq.~\eqref{4}, leads to
\begin{equation}
 \frac{\partial\Phi}{\partial\tilde{\phi}_1}\Bigg\vert_{J_1}
 = \left(J_2-\hbar j_2\right)
 \frac{\partial\tilde{\phi}_2}{\partial\tilde{\phi}_1}\Bigg\vert_{J_1}
 \label{4e}
\end{equation}
and
\begin{equation}
\frac{\partial^2\Phi}{\partial\tilde{\phi}_1^2}\Bigg\vert_{J_1}
 = \frac{\partial J_2}{\partial\tilde{\phi}_1}\Bigg\vert_{J_1}\frac{\partial\tilde{\phi}_2}{\partial\tilde{\phi}_1}\Bigg\vert_{J_1}
 +\left(J_2-\hbar j_2\right)
 \frac{\partial^2\tilde{\phi}_2}{\partial\tilde{\phi}_1^2}\Bigg\vert_{J_1}. 
 \label{4e}
 \end{equation}
Therefore, the values of $\tilde{\phi}_1$ making $\Phi$ stationary are those leading to
\begin{equation}
 J_2=\hbar j_2 
 \label{4f}
\end{equation}
or
\begin{equation}
 \frac{\partial\tilde{\phi}_2}{\partial\tilde{\phi}_1}\Bigg\vert_{J_1}=0. 
 \label{4g}
\end{equation}
From Eqs.~\eqref{5}, \eqref{8} and~\eqref{4e}, we finally arrive at
\begin{equation}
 S_{j_2j_1}(E) = \sum_{k} 
 \;
 \left(
 \frac{2\pi i}{\hbar}
 \left|
 \frac{\partial J_2}{\partial\tilde{\phi}_1}\Bigg\vert_{J_1}^k
 \right|
 \right)^{-1/2}
 e^{i\left(\Omega_k/\hbar-\pi(\nu_k+\chi_k)/2
 \right)}.
 \label{9}
 \end{equation}
The sum is over those trajectories satisfying Eq.~\eqref{4f}. Since the prefactor of the integrand in Eq.~\eqref{5} is 
$\big\vert\frac{\partial\tilde{\phi}_2}{\partial\tilde{\phi}_1}\big\vert_{J_1}\big\vert^{1/2}$, it is clear from Eqs.~\eqref{5}, \eqref{8} 
and~\eqref{4e} that the contribution to $S_{j_2j_1}(E)$ of those trajectories complying with Eq.~\eqref{4g} is zero. 
We thus recover Eq.~\eqref{7}, but with the index $\eta_k$ replaced by $\nu_k+\chi_k$. As previously stated, the RHS of Eqs.~\eqref{7}
and~\eqref{9} may differ by an irrelevant phase factor $e^{i\pi l/2}$. Multiplying the latter by the RHS of Eq.~\eqref{7} and equating 
the resulting product with the RHS of Eq.~\eqref{9} leads to the condition for the validity of Eq.~\eqref{5}, namely
\begin{equation}
 \eta_k=\nu_k+\chi_k+l.
 \label{10}
 \end{equation}
For clarity's sake, we recall that (i) $\eta_k$ is the number of caustics touched by the trajectories, (ii) $\nu_k$ is 0 if 
$\frac{\partial\tilde{\phi}_2}{\partial\tilde{\phi}_1}\big\vert_{J_1}$ is positive, $\pm$1 otherwise, and (iii) $\chi_k$ is 0 if 
$\frac{\partial J_2}{\partial\tilde{\phi}_1}\big\vert_{J_1}\frac{\partial\tilde{\phi}_2}{\partial\tilde{\phi}_1}\big\vert_{J_1}$ is positive, 
1 otherwise. $l$ is thus necessarily an integer.

\begin{figure}[!h]

\centering{\includegraphics[width=27cm]{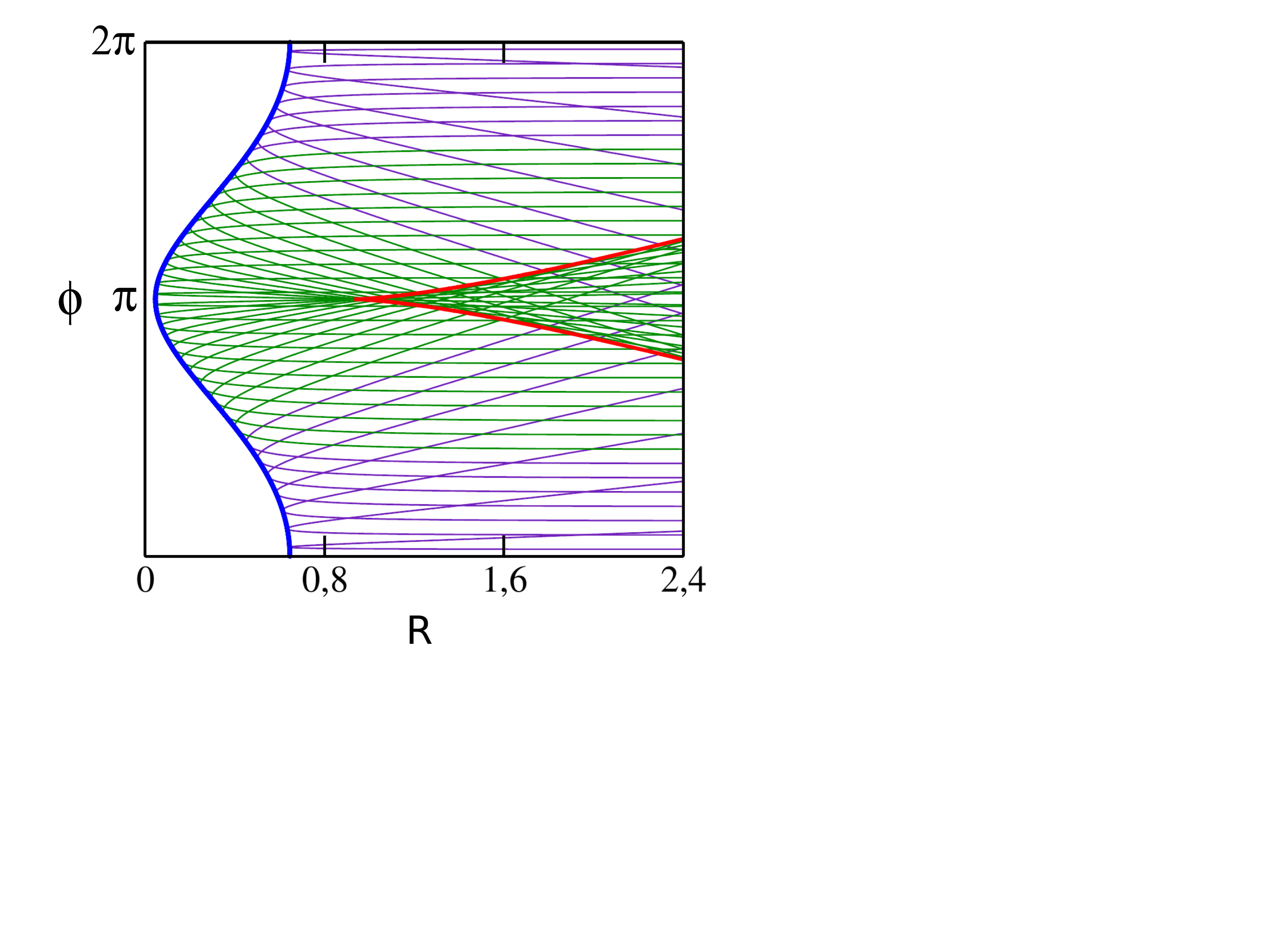}}

\caption{Set of 36 trajectories for $\beta=0.3$. Indigo paths touch only one caustic, represented in blue, while green paths touch
both the previous caustic and a second one represented in red (see text for more details).}

\label{fig:3}

\end{figure}

We now want to know when Eq.~\eqref{10} is satisfied. To answer this question, 36 trajectories projected onto the $(R,\phi)$ plane are represented
in Fig.~\ref{fig:3} for $\beta=0.3$. Trajectories initially come from the right with $J_1=0$ and are thus parallel to the $R$-axis. 
They rebound against the anisotropic 
potential wall which rotationally excites the diatom (except for $\phi_1=0$ and $\pi$). The final directions of the rotationally 
excited trajectories make a non zero angle with the $R$-axis. All the trajectories, indigo plus green, touch a first caustic, represented
by a blue thick line lying within the interaction region. On the other hand, only the green paths touch a second caustic, represented by a 
red thick line. $\eta_k$ is thus equal to 1 for the indigo paths, and 2 for the green paths. Moreover, we have previously seen that 
$\frac{\partial\tilde{\phi}_2}{\partial\tilde{\phi}_1}\big\vert_{J_1}$ is always positive for $\beta=0.3$ (see Fig.~\ref{fig:2}), thus implying
$\nu_k=0$. Hence, we deduce from the definition of $\chi_k$ (see above) that $\chi_k=0$ if 
$\frac{\partial J_2}{\partial\tilde{\phi}_1}\big\vert_{J_1}$ is positive, 1 otherwise. Now, we have seen previously that 
$\frac{\partial J_2}{\partial\tilde{\phi}_1}\big\vert_{J_1}$ is positive (negative) for one (two) caustic(s). 
Consequently, $\chi_k = 0$ when $\eta_k = 1$ while $\chi_k = 1$ when $\eta_k = 2$. Eq.~\eqref{10} 
is thus satisfied whatever $\tilde{\phi}_1$ if $l$ is taken at 1. This scenario holds for $\beta=0.1$.

On the other hand, the situation is more complex for $\beta=1.02$ as trajectories touch up to three caustics. 
The difference between $\eta_k$ and $\nu_k+\chi_k$  is found to
strongly vary in terms of $\tilde{\phi}_1$ for both definitions of $\nu_k$ [(0,+1) or (0,-1)]. 
As a consequence, Eq.~\eqref{10} cannot be satisfied whatever $\tilde{\phi}_1$ with a single value of $l$. 
A sufficiently small coupling between translational and rotational 
motions within the interaction region appears to be a prerequisite for the validity of Eq.~\eqref{5}.

\section{Alternative phase index}
\label{IV}

\begin{figure}[!h]

\centering{\includegraphics[width=27cm]{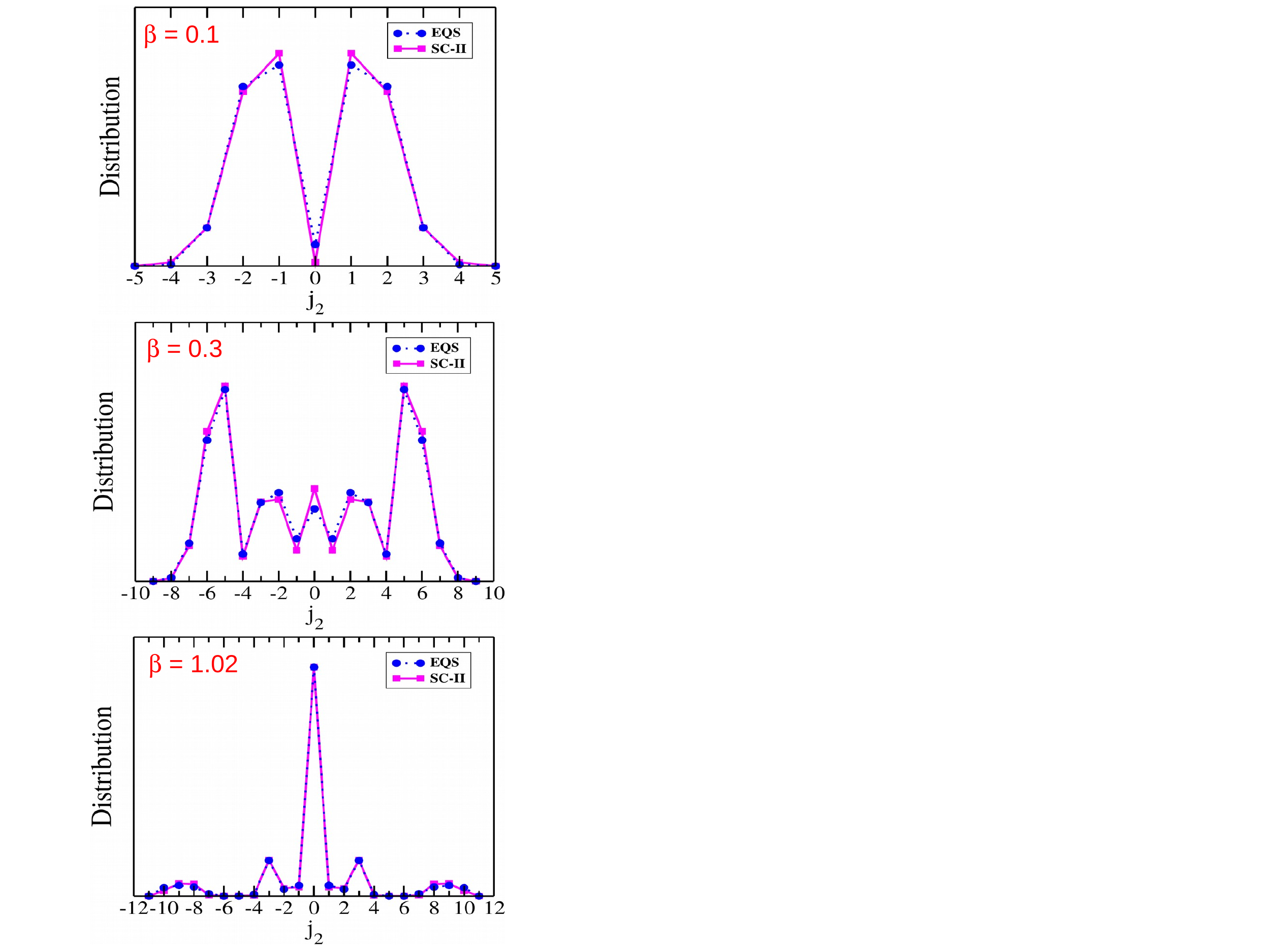}}

\caption{Rotational state distributions found from exact quantum scattering (EQS) calculations (blue circles connected by dotted segments)
and semiclassical (SC-II) calculations according to Eq.~\eqref{12} (magenta squares connected by solid segments).}

\label{fig:5}

\end{figure}

Eq.~\eqref{10} suggests the expression of the index making general Eq.~\eqref{5}. It is indeed sufficient to replace 
$\nu$ by $\eta-\chi-l$ in Eq.~\eqref{5}. Thanks to this substitution, the latter will necessarily lead to Eq.~\eqref{7} within the SPA.
Related developments have been performed by Campolieti and Brumer in order to derive phase indices for various SCIVR time-dependent 
propagators.\cite{Campo2} 
Since only the relative phases of $S$-matrix elements are relevant, the value of $l$ is irrelevant. Taking it at 0 leads to
\begin{equation}
 S_{j_2j_1}(E) = \frac{1}{2\pi i} \int_{0}^{2\pi} d\tilde{\phi}_1 
 \;\left|
 \frac{\partial\tilde{\phi}_2}{\partial\tilde{\phi}_1}\Bigg\vert_{J_1}
 \right|^{1/2}
 e^{i\left(\Phi/\hbar-\pi(\eta-\chi)/2\right)}.
 \label{11}
\end{equation}
A slightly different alternative to this expression, suggested by previous developments,\cite{BC1,BC2} is
\begin{equation}
 S_{j_2j_1}(E) = \frac{1}{2\pi i} \int_{0}^{2\pi} d\tilde{\phi}_1 
 \;\left|
 \frac{\pi_2}{P_2}
 \frac{\partial\tilde{\phi}_2}{\partial\tilde{\phi}_1}\Bigg\vert_{J_1}
 \right|^{1/2}
 e^{i\left(\Phi/\hbar-\pi(\eta-\chi)/2\right)},
 \label{12}
\end{equation}
where $\pi_2$ is given by Eq.~\eqref{3b} with $J_2=\hbar j_2$. The advantage of Eq.~\eqref{12} over Eq.~\eqref{11} is that 
$S_{j_2j_1}(E)$ is rigorously zero for energetically prohibited transitions. Within the SPA, however, both expressions lead 
to Eq.~\eqref{7}.
The rotational state distributions obtained by means of Eq.~\eqref{12}, labeled SC-II, are compared with EQS distributions in 
Fig.~\ref{fig:5}. For $\beta$ equal 0.1 and 0.3, things are obviously unchanged with respect to Fig.~\ref{fig:1}, but for $\beta$ equal 1.02, 
the agreement between semiclassical and quantum calculations is considerably improved. Eq.~\eqref{11} leads to virtually identical results.

\section{Discussion}
\label{V}

\begin{figure}[!h]

\centering{\includegraphics[width=18cm]{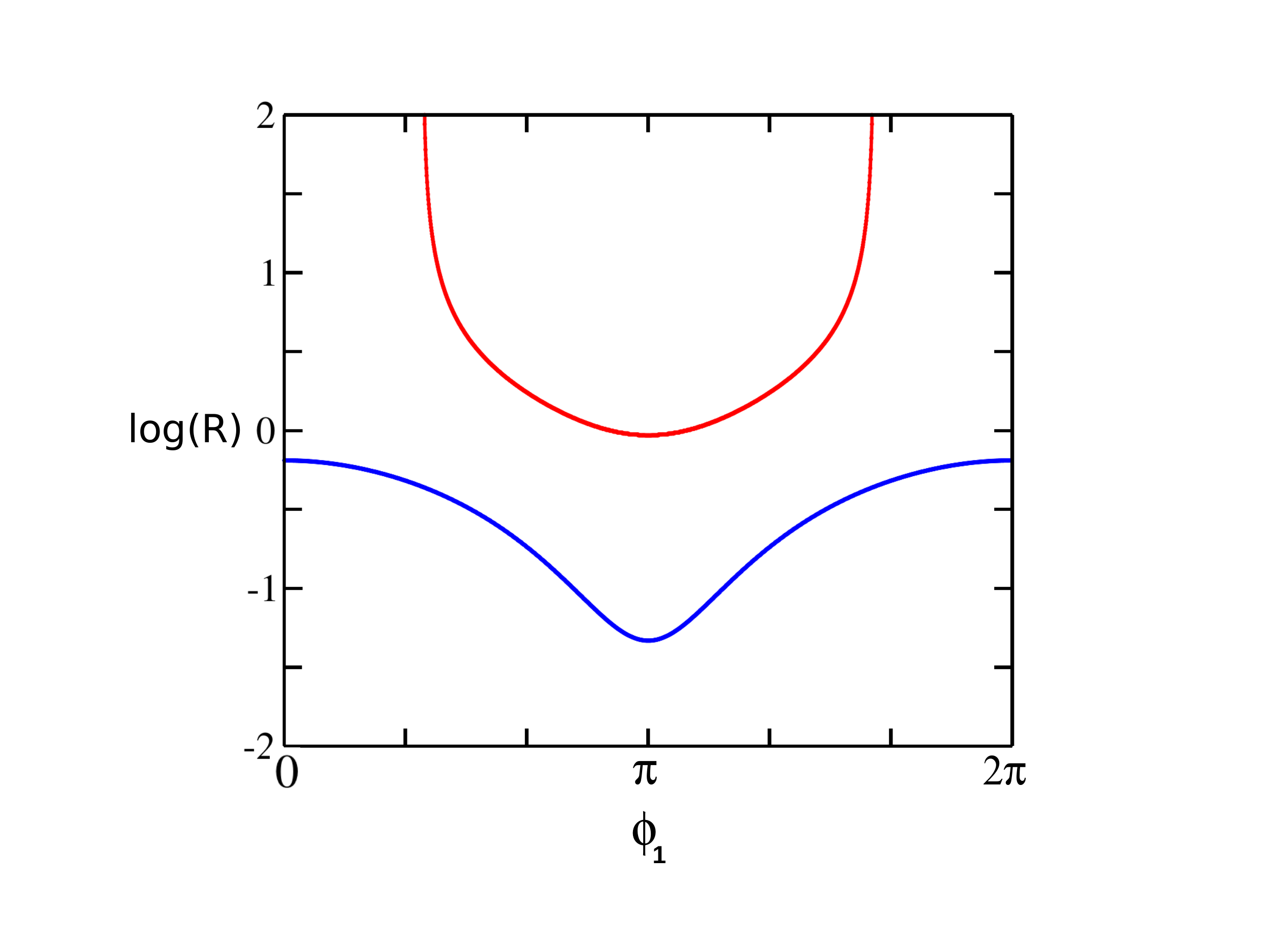}}

\caption{Values of $log(R)$ in terms of $\phi_1$ at the blue and red caustics (see Fig.~\ref{fig:3} for their representation in the ($R,\phi$)
plane). }

\label{fig:6}

\end{figure}

\begin{figure}[!h]

\centering{\includegraphics[width=18cm]{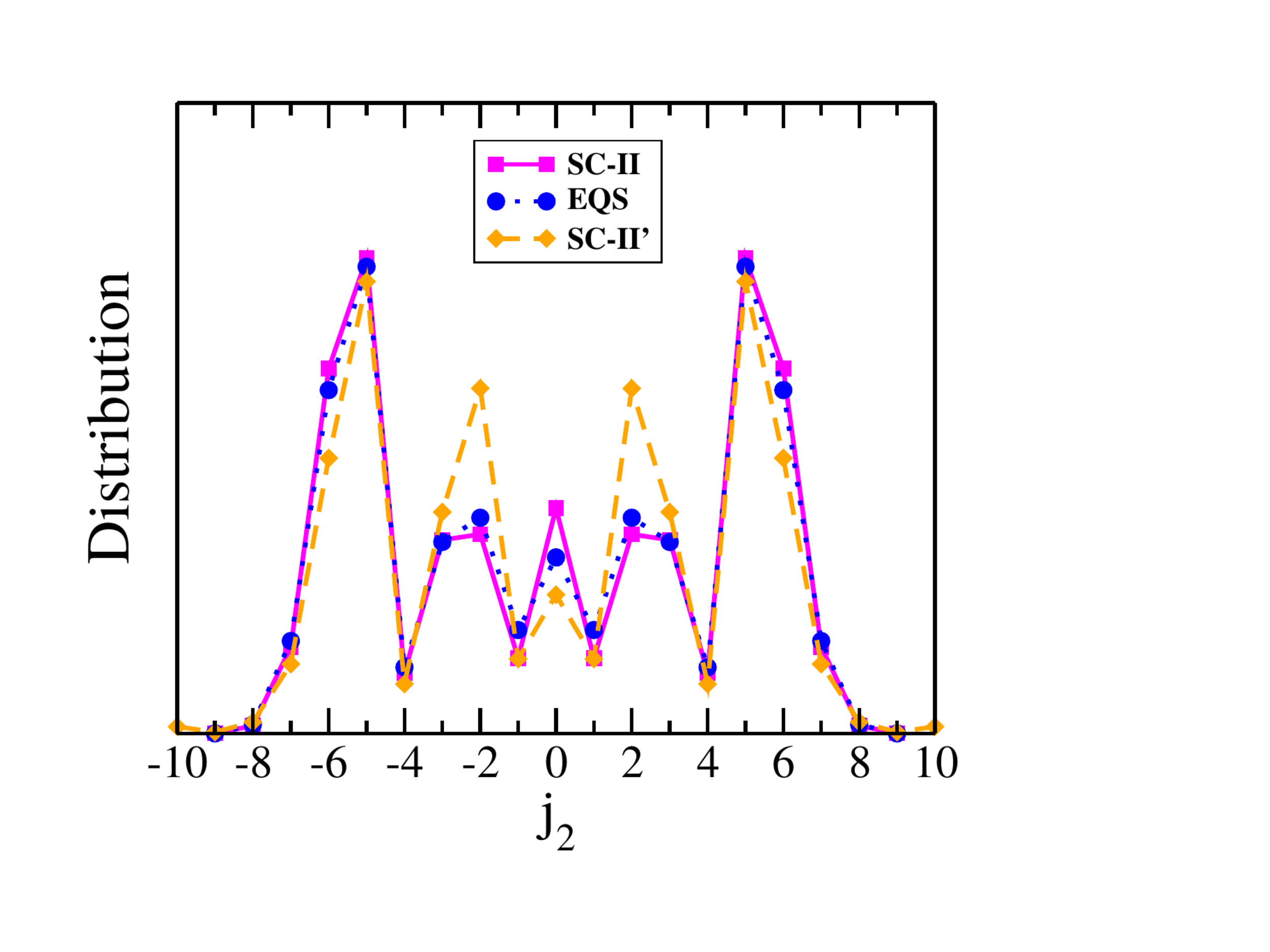}}

\caption{Rotational state distributions found from exact quantum scattering (EQS) calculations (blue circles connected by dotted segments),
converged semiclassical (SC-II) calculations according to Eq.~\eqref{12} for $R_2=10^3$ \AA\;(magenta squares connected by solid segments),
and non converged semiclassical (SC-II') calculations according to Eq.~\eqref{12} for $R_2=10$ \AA\;(orange diamonds connected by dashed segments).}

\label{fig:7}

\end{figure}

The \emph{sine qua non} condition for obtaining accurate semiclassical predictions from Eqs.~\eqref{11} or~\eqref{12} is that $R_2$ is taken 
in principle at infinity, in practice at a large value (at least $\sim10^2$ \AA). The reason is as follows. For $\beta=0.3$, 
the value of $log(R)$ at the caustics is 
represented in terms of $\phi_1$ in Fig.~\ref{fig:6}. The blue and red curves correspond, respectively, to the blue and red caustics in 
Fig.~\ref{fig:3}. If one takes $R_2$ at $\sim$10 \AA, as is commonly done in classical trajectory calculations, the Maslov index $\eta$ is 
found equal to 1 for all the trajectories crossing the red caustic beyond $\sim$10 \AA, i.e., for all the paths such that $log(R)$ is 
larger than $\sim$1 (see Fig.~\ref{fig:6}). This is a wrong estimation since $\eta=2$ for these paths. Though the latter represent a small 
percentage of the whole 
set of trajectories contributing to $S$-matrix elements (see Fig.~\ref{fig:6}), the alteration of the rotational state distribution is 
significant, as seen in Fig.~\ref{fig:7}. Total disagreement is found for $R_2=4$ \AA, value for which the purely classical predictions, 
or the semiclassical ones according to Eq.~\eqref{5}, are already converged. To get the red curve in Fig.~\ref{fig:6}, it was necessary 
to take $R_2$ at $10^2$ \AA. For obtaining the rotational distributions, however, such calculations are not only heavy, they are useless. 
It is sufficient to run trajectories up to $4$ \AA, and then to analytically deduce the trajectory conditions at any larger value 
of $R_2$ (we took it at $10^3$) from those at 4 \AA. On the other hand, there is no need to take $R_1$ at a large value, since the first caustic (blue line in 
Fig.~\ref{fig:3}) lies within the interaction region. $R_1$ was thus taken at 4 \AA\;for all the calculations related to this work. 

The value of the norm [$\Sigma=\Sigma_{j_2}P_{j_2j_1}(E)$] is an excellent criterion of accuracy. The closer to 1, the more accurate the 
semiclassical predictions. In a first series of calculations, we found that for 1800 trajectories, Eq.~\eqref{12} leads to $\Sigma=0.9999$ 
for $\beta=0.1$ (only the significant digits are given), $\Sigma=0.999$ for $\beta=0.3$ and $\Sigma=1.02$ for $\beta=1.02$.
Eq.~\eqref{11} leads to nearly identical predictions. One notes that the 
accuracy decreases with the coupling strength between the $R$ and $\phi$ coordinates, i.e., when trajectories become more and more unstable. 
In a second series of calculations, we sought to determine the minimum number of trajectories maintaining the previous accuracy. For 
$\beta=0.1$ and 0.3, we arrived at the amazingly low numbers of 16 and 30, respectively. These numbers are of the order of those found in 
previous studies.\cite{McCurdy,ElranKay} This is far less than the number of trajectories 
necessary to estimate final state distributions by means of the classical trajectory method which totally misses interferences ! 
On the other hand, a much larger number of 660 trajectories is mandatory for $\beta=1.02$. When $\beta$ increases, calculations are
more time-consuming in addition to be less accurate. 

The fact that a few tens of trajectories may be enough for accurate semiclassical predictions on inelastic collisions 
involving one rotational degree-of-freedom (DOF) augurs well for the extension of the present work to realistic atom-rigid diatom 
collisions involving two rotational DOFs. For such processes, Miller's SCIVR $S$-matrix elements are indeed given by integrals over the two 
angles conjugate to $j$ and $l$, respectively.\cite{Campo1} For nearly isotropic potential energy surfaces such as those involved in 
stereodynamics studies, and for given initial values ($j_1,l_1)$ of $(j,l)$, $\sim10^3$ trajectories (the square of 30) should be enough for 
converging the calculation of $S^J_{j_2l_2,j_1l_1}$, where $J$ is now the total angular momentum, and $j_2$ and $l_2$ can take any values consistent 
with the conservation of $E$ and $J$. For $j_1=0$, a few tens of thousands of trajectories are expected to be sufficient to obtain the ICS, and 
hence, the steric asymmetry. 
Note that the analytical extrapolation of the dynamics from a few \AA\;to infinity is feasible for an atom-rigid diatom system. 

A few years ago, several SCIVR approaches of rotational transitions were proposed within the standard configuration space coordinates $(R,\phi)$ 
\cite{BC1,BC2}. These formulations all involve phase indices making them in principle of general applicability. Unfortunately, however, 
they also involve integrals of functions oscillating all the more as $R_1$ and $R_2$ take large values, making thereby challenging 
the numerical convergence of $S$-matrix elements in the asymptotic channel. Eqs.~\eqref{11} and~\eqref{12} do not suffer from this drawback,
for their integrands do not depend on $R_1$ and $R_2$ in the asymptotic channel (apart from $\eta$ which we know how to calculate).

Amazingly accurate SCIVR calculations have been performed by Elran and Kay.\cite{ElranKay} To date, however, their approach has only been applied 
to collinear processes, and its applicability to three-dimensional collisions is an open issue.\cite{ElranKay} Moreover, the mathematical form of their $S$-matrix 
elements is much more complex than the one of Eqs.~\eqref{11} and~\eqref{12}. It is thus unclear whether this approach can be used to explain quantum 
interferences as efficiently as Eqs.~\eqref{11} or~\eqref{12}. This is all the more so as the previous equations already provide 
quasi-quantitative results (see Fig.~\ref{fig:5}). 
 

 


\section{Conclusion}
\label{VI}

Miller's SCIVR theory in the interaction picture\cite{McCurdy} (Eq.~\eqref{5}) was presented and applied to a model of atom-planar rotor 
inelastic collision involving strong quantum interferences. Three coupling strengths between translational and rotational motions were considered. 
For the two lowest ones, SCIVR predictions were found to be in close agreement with quantum scattering results. For the strongest one, however, 
clear disagreement was observed. In order to shed light on this finding, the conditions of validity of Eq.~\eqref{5} were analyzed.
We found that the latter tends to classical $S$-matrix theory\cite{Bill1,Bill2,Marcus} in the classical limit only whether the coupling strength 
is sufficiently small. This inconsistency was removed by substituting a new phase index to the original one. SCIVR predictions were 
then found to be in close agreement with quantum scattering results whatever the coupling strength. The next step will be to apply
the present work to the realistic case of three-dimensional rotationally inelastic collisions. In the event that the resulting approach 
provides accurate predictions, we will have a powerful tool at our disposal to better analyze rotational state distributions of astrochemical 
interest,\cite{Lique} or the state-of-the-art stereodynamics measurements currently performed.\cite{Brouard0,Groenenboom,Brouard1,Brouard2}

\bibliography{Biblio_TOT}

\end{document}